\documentclass[final,5p,times,twocolumn]{elsarticle}
\usepackage{lineno,hyperref,amsmath,amsthm,amssymb,amsfonts,ragged2e,color,subfig}
\journal{``The European Physical Journal D"}
\begin{document}
\begin{frontmatter}
\title{Dust-acoustic rogue waves in an electron depleted plasma}
\author{R. K. Shikha$^{a,1}$, N. A. Chowdhury$^{b,1}$, A. Mannan$^{1,2}$, and A. A. Mamun$^{1,3}$}
\address{$^{1}$Department of Physics, Jahangirnagar University, Savar, Dhaka-1342, Bangladesh\\
$^{2}$Institut f\"{u}r Mathematik, Martin Luther Universit\"{a}t Halle-Wittenberg, Halle, Germany\\
$^{3}$Wazed Miah Science Research Center, Jahangirnagar University, Savar, Dhaka-1342, Bangladesh\\
Email: $^*$shikha261phy@gmail.com, $^*$nurealam1743phy@gmail.com}
\begin{abstract}
A rigorous theoretical investigation is made to study the characteristics of dust-acoustic (DA)
waves (DAWs) in an electron depleted unmagnetized opposite polarity dusty plasma system that contains
super-thermal ($\kappa$-distributed) ions, mobile positively and negatively charged dust grains for the
first time. The reductive perturbation method is employed to obtain the NLSE to explore the
modulational instability (MI) conditions for DAWs as well as the formation and characteristics of
gigantic rogue waves. The nonlinear and dispersion properties of the dusty plasma medium are the
prime reasons behind the formation of rogue waves. The height and thickness of the DARWs associated
with DAWs as well as the MI conditions of DAWs are numerically analyzed by changing different dusty
plasma parameters, such as dust charges, dust and ion number densities, and ion-temperature, etc.
The implications of the results for various space dusty plasma systems (viz.,  mesosphere, F-rings of Saturn, and
cometary atmosphere, etc.) as well as laboratory dusty plasma produced by laser-matter interaction are briefly mentioned.
\end{abstract}
\end{frontmatter}
\section{Introduction}
\label{1sec:Introduction}
Opposite polarity (OP) dusty plasma (OPDP) is characterised as fully ionized gas,
comprising massive positively and negatively charged dust grains as well as electrons
and ions, and is believed to exist in space, viz., Planetary rings \cite{Shukla1992},
Jupiter's magnetosphere \cite{Hossen2016a}, interstellar clouds \cite{Hossen2016b,Hossen2017,Shahmansouri2013}, Earth
polar mesosphere \cite{Hossen2016a}, cometary tails \cite{Hossen2016a}, solar system
\cite{C1,C2,C3} and laboratory situations, viz., laser-matter interaction
\cite{Shahmansouri2013}. Rao \textit{et al.} \cite{Rao1990} have first theoretically
predicted a new kind of low-frequency dust-acoustic (DA) waves (DAWs), and this
low-frequency DAWs have been further experimentally identified  by Barkan
\textit{et al.} \cite{Barkan1995} in dusty plasma (DP) medium (DPM). A revolution associated with DP physics has
been welcomed after experimental identification of the DAWs, and many researchers have
performed various modern eigen modes, viz., DAWs \cite{Hossen2016a,Hossen2016b,Hossen2017},
dust lattice waves \cite{Melandso1996}, dust-drift waves \cite{Shukla1991},
DA shock waves (DASHWs) \cite{Ferdousi2015}, DA solitary waves (DASWs) \cite{Shahmansouri2013} and
dust-ion-acoustic waves (DIAWs) \cite{Shukla1992} in DPM to understand various nonlinear structures
regarding the propagation of low frequency electrostatic perturbation.

The attachment of electrons with massive dust grains from the ambient DPM during the dust charging process
is referred to as electron depletion \cite{Ferdousi2015,Mamun1996,Dialynas2009,Mayout2012,Sahu2012,Ferdousi2017}.
The signature of electron depletion mechanism, in which majority even sometimes all the electrons are inserted into
the massive dust grains, associated to the dust can be observed in space environments, viz., F-rings of
Saturn \cite{Mayout2012}, Jupiter's magnetosphere \cite{Hossen2016a},
interstellar clouds \cite{Hossen2017}, Earth polar mesosphere \cite{Hossen2016a},
cometary tails \cite{Hossen2016a}, solar system \cite{Hossen2017}, and laboratory DPM.
Shukla and Silin studied DIAWs in an electron depleted DPM (EDDPM). Mamun \textit{et al.} \cite{Mamun1996}
examined solitary potentials in two components EDDPM, and found that both dust
and ion densities enhance the negative potentials. Sahu and Tribeche \cite{Sahu2012} reported the
small amplitude double-layers (DLs) in an unmagnetized EDDPM, and demonstrated that their model can admit both
compressive and rarefactive DA DLs (DADLs) according to the properties of plasma parameters.
Ferdousi \textit{et al.} \cite{Ferdousi2015} studied DASHWs in two components EDDPM,
and found that under consideration, their model supports both positive and negative potentials.
Hossen \textit{et al.} \cite{Hossen2016a,Hossen2016b} investigated DAWs in three components EDDPM
having inertial massive OP dust grains (OPDGs) and inertialess non-thermal ions, and observed that the
presence of the positively charged dust significantly modified the shape of DASWs
and DADLs potential structures.

The super-thermal or $\kappa$-distribution \cite{Vasyliunas1968,C4,C5,Shahmansouri2012,Kourakis2011,Uddin2015,C6,Sultana2011,Ahmed2018,Gill2010}
can describe the deviation, according to the values of the super-thermal parameter
$\kappa$ which manifests the presence of the external force fields or wave-particle
interaction, of plasma species from the thermal or Maxwellian distribution. The super-thermal or
$\kappa$-distribution exchanges with the Maxwellian distribution when $\kappa$ tends
to infinity, i.e., $\kappa\rightarrow\infty$, and $\kappa$-distribution
is normalizable for any kind of values of $\kappa$ by fulfilling this
condition $\kappa>3/2$ \cite{Shahmansouri2012,Kourakis2011,Uddin2015,Sultana2011,Ahmed2018,Gill2010}.
Shahmansouri and Alinejad \cite{Shahmansouri2012} investigated DASWs in a super-thermal DPM,
and found that the depth of the potential well decreases with increasing the value of $\kappa$.
Kourakis and Sultana \cite{Kourakis2011} examined  the presence of the super-thermal
particles in a DPM, and observed how the fast particles change the speed of the DIA solitons, and also
found that lower $\kappa$ values support faster solitons. Uddin \textit{et al.} \cite{Uddin2015}
analyzed the nonlinear propagation of positron-acoustic waves in a super-thermal plasma, and highlighted that the height
of the positive potential decreases with increasing value of $\kappa$.

The modulational instability (MI), energy localization, and energy redistribution of the carrier
waves are governed by the standard nonlinear Schr\"{o}dinger equation (NLSE) \cite{Sultana2011,Ahmed2018,Gill2010,C7,Saini2018,C8,Kourakis2005}.
Sultana and Kourakis \cite{Sultana2011} studied electron-acoustic (EA) envelope solitons in presence of
super-thermal electrons, and observed that the unstable domain of EA waves increases with $\kappa$.
Ahmed \textit{et al.} \cite{Ahmed2018} examined ion-acoustic waves in multi-component plasmas, and
demonstrated that the critical wave number ($k_c$) decreases with the increase of $\kappa$.
Gill \textit{et al.} \cite{Gill2010} investigated the MI of the DAWs in presence of super-thermal
ions in a DPM, and found that the excess super-thermality of the ions enhances the stable domain of the DAWs.
Saini and Kourakis \cite{Saini2018} reported amplitude modulation of the DAWs in presence of the super-thermal
ions in a DPM, and  the excess super-thermality of the plasma species recognizes narrower envelope solitons.
Kourakis and Shukla \cite{Kourakis2005} demonstrated the MI of the DAWs in an OPDP.

Recently, Shahmansouri and Alinejad \cite{Shahmansouri2013} demonstrated DASWs in an EDDPM in presence
of super-thermal plasma species, and found that the height of the DASWs increases with
the increase in the value of super-thermality of plasma particles. In this paper, we want to develop sufficient
extension of previous published work \cite{Shahmansouri2013} by presenting a real and novel three
component DP model. It could be of interest to examine the MI of DAWs and formation of DA rogue
waves (DARWs) a by considering a three component DP model having highly charged massive
OPDGs as well as inertialess ions are modelled by the super-thermal $\kappa$-distribution.
\section{Model Equations}
\label{1:Model Equations}
We consider a three component unmagnetized EDDPM  comprising inertial
negatively and positively charged massive dust grains, and $\kappa$-distributed positive ions.
At equilibrium, the quasi-neutrality condition can be written as
$Z_in_{i0}+Z_+n_{+0}\approx Z_- n_{-0}$; where $n_{i0}$, $n_{-0}$, and $n_{+0}$ are
the number densities of positive ions, negative and positive dust grains, respectively,
and $Z_i$, $Z_+$ and $Z_-$ are the charge state of the positive ion, positive and negative
dust grains, respectively. So, the normalizing equations to study the DAWs are
\begin{eqnarray}
&&\hspace*{-1.3cm}\frac{\partial n_+}{\partial t}+\frac{\partial}{\partial x}(n_+u_+)=0,
\label{1:eq1}\\
&&\hspace*{-1.3cm}\frac{\partial u_+}{\partial t}+u_+\frac{\partial u_+}{\partial x}=-\frac{\partial \phi}{\partial x},
\label{1:eq2}\\
&&\hspace*{-1.3cm}\frac{\partial n_-}{\partial t}+\frac{\partial}{\partial x}(n_-u_-)=0,
\label{1:eq3}\\
&&\hspace*{-1.3cm}\frac{\partial u_-}{\partial t}+u_-\frac{\partial u_-}{\partial x}=s_1\frac{\partial \phi}{\partial x},
\label{1:eq4}\\
&&\hspace*{-1.3cm}\frac{\partial^2 \phi}{\partial x^2}=s_2n_--(s_2-1)n_i-n_+,
\label{1:eq5}\
\end{eqnarray}
where $n_i$, $n_-$, and $n_+$ are normalized by $n_{i0}$, $n_{-0}$, and $n_{+0}$, respectively;
$u_+$ and $u_-$ represent the positive and negative dust fluid speed, respectively, normalized
by the DA wave speed $C_{+}=(Z_+k_BT_i/m_+)^{1/2}$ (with $T_i$ being
temperature of ion, $m_+$ being positive dust mass, and $k_B$ being the Boltzmann constant); $\phi$
represents the electrostatic wave potential normalized by $k_BT_i/e$ (with $e$ being the magnitude
of single electron charge); the time and space variables are, respectively, normalized by
$\omega_{P_+}^{-1}=(m_+/4\pi e^{2}Z_+^{2}n_{+0})^{1/2}$, and $\lambda_{D_+} = (k_BT_i/4\pi e^2Z_+n_{+0})^{1/2}$.
Other parameters can be defined as $s_1=Z_-m_+/Z_+m_-$ and $s_2=Z_-n_{-0}/Z_+n_{+0}$. It may be
noted here that we have considered $m_->m_+$, $Z_->Z_+$, and $n_{-0}>n_{+0}$.
The expression for the number density of ions following the $\kappa$-distribution
\cite{Shahmansouri2013} can be written as
\begin{eqnarray}
&&\hspace*{-1.3cm}n_i=\left[1 +\frac{\phi}{(\kappa-3/2)}\right]^{-\kappa+\frac{1}{2}}
\label{1:eq6}\
\end{eqnarray}
where the parameter $\kappa$ is known as  super-thermality of the ions.
Now, by substituting Eq. \eqref{1:eq6} into Eq. \eqref{1:eq5}, and expanding up to third order in $\phi$, we obtain
\begin{eqnarray}
&&\hspace*{-1.3cm}\frac{\partial^2 \phi}{\partial x^2}+n_+-s_2n_-= (1-s_2)+M_1\phi
\nonumber\\
&&\hspace*{+1.2cm}+M_2\phi^2+M_3\phi^3+\cdot\cdot\cdot,
\label{1:eq7}\
\end{eqnarray}
where
\begin{eqnarray}
&&\hspace*{-1.3cm}M_1 = \frac{(s_2-1)(2\kappa-1)}{(2\kappa-3)},
\nonumber\\
&&\hspace*{-1.3cm}M_2 =\frac{(1-s_2)(2\kappa-1)(2\kappa+1)}{2(2\kappa-3)^2},
\nonumber\\
&&\hspace*{-1.3cm}M_3 =\frac{(s_2-1)(2\kappa-1)(2\kappa+1)(2\kappa+3)}{6(2\kappa-3)^3}.
\nonumber\
\end{eqnarray}
We note that the term on the right hand side is the contribution of positive ions.
\section{Derivation of the NLSE}
\label{1:Derivation of the NLSE}
To study the MI of DAWs, we will derive the NLSE by employing the
reductive perturbation method. So, we first introduce the
stretched co-ordinates
\begin{eqnarray}
&&\hspace*{-1.3cm}\xi={\epsilon}(x-v_g t),
\label{1:eq8}\\
&&\hspace*{-1.3cm}\tau={\epsilon}^2 t,
\label{1:eq9}\
\end{eqnarray}
where $v_g$ is the group speed and $\epsilon$ is a small parameter. Then, we can write the dependent variables as
\begin{eqnarray}
&&\hspace*{-1.3cm}n_+=1+\sum_{m=1}^{\infty}\epsilon^{m}\sum_{l=-\infty}^{\infty}n_{+l}^{(m)}(\xi,\tau)\mbox{exp}[i l(kx-\omega t)],
\label{1:eq10}\\
&&\hspace*{-1.3cm}u_+=\sum_{m=1}^{\infty}\epsilon^{m}\sum_{l=-\infty}^{\infty}u_{+l}^{(m)}(\xi,\tau)\mbox{exp}[i l(kx-\omega t)],
\label{1:eq11}\\
&&\hspace*{-1.3cm}n_-=1+\sum_{m=1}^{\infty}\epsilon^{m}\sum_{l=-\infty}^{\infty}n_{-l}^{(m)}(\xi,\tau)\mbox{exp}[i l(kx-\omega t)],
\label{1:eq12}\\
&&\hspace*{-1.3cm}u_-=\sum_{m=1}^{\infty}\epsilon^{m}\sum_{l=-\infty}^{\infty}u_{-l}^{(m)}(\xi,\tau)\mbox{exp}[i l(kx-\omega t)],
\label{1:eq13}\\
&&\hspace*{-1.3cm}\phi=\sum_{m=1}^{\infty}\epsilon^{m}\sum_{l=-\infty}^{\infty}\phi_l^{(m)}(\xi,\tau)\mbox{exp}[i l(kx-\omega t)].
\label{1:eq14}\
\end{eqnarray}
where $k$ ($\omega$) is real variables representing the carrier wave number (frequency). The derivative operators in the above equations are treated as follows:
\begin{eqnarray}
&&\hspace*{-1.3cm}\frac{\partial}{\partial t}\rightarrow\frac{\partial}{\partial t}-\epsilon v_g\frac{\partial}{\partial\xi}+\epsilon^2\frac{\partial}{\partial\tau},
\label{1:eq15}\\
&&\hspace*{-1.3cm}\frac{\partial}{\partial x}\rightarrow\frac{\partial}{\partial x}+\epsilon\frac{\partial}{\partial\xi}.
\label{1:eq16}\
\end{eqnarray}
Now, by substituting Eqs. \eqref{1:eq10}$-$\eqref{1:eq16} into Eqs. \eqref{1:eq1}$-$\eqref{1:eq4} and Eq. \eqref{1:eq7}, and collecting
the terms containing $\epsilon$, the first order ($m=1$ with $l=1$) equations can be expressed as
\begin{eqnarray}
&&\hspace*{-1.3cm}\omega n_{+1}^{(1)}=ku_{+1}^{(1)},
\label{1:eq17}\\
&&\hspace*{-1.3cm}k\phi_1^{(1)}=\omega u_{+1}^{(1)},
\label{1:eq18}\\
&&\hspace*{-1.3cm}\omega n_{-1}^{(1)}=ku_{-1}^{(1)},
\label{1:eq19}\\
&&\hspace*{-1.3cm}ks_1\phi_1^{(1)}=-\omega u_{-1}^{(1)},
\label{1:eq20}\\
&&\hspace*{-1.3cm}n_{+1}^{(1)}=k^{2}\phi_1^{(1)}+M_1\phi_1^{(1)}+s_2n_{-1}^{(1)},
\label{1:eq21}\
\end{eqnarray}
these equations reduce to
\begin{eqnarray}
&&\hspace*{-1.3cm}n_{+1}^{(1)}=\frac{k^2}{\omega^2}\phi_1^{(1)},
\label{1:eq22}\\
&&\hspace*{-1.3cm}u_{+1}^{(1)}=\frac{k}{\omega}\phi_1^{(1)},
\label{1:eq23}\\
&&\hspace*{-1.3cm}n_{-1}^{(1)}=-\frac{s_1k^2}{\omega^2}\phi_1^{(1)},
\label{1:eq24}\\
&&\hspace*{-1.3cm}u_{-1}^{(1)}=-\frac{k s_1}{\omega}\phi_1^{(1)},
\label{1:eq25}\
\end{eqnarray}
we thus obtain the dispersion relation for DAWs
\begin{eqnarray}
&&\hspace*{-1.3cm}\omega^{2}=\frac{k^{2}(1+s_1 s_2)}{M_1+k^{2}}\Bigg.
\label{1:eq26}\
\end{eqnarray}
The second-order ($m=2$ with $l=1$) equations are given by
\begin{eqnarray}
&&\hspace*{-1.3cm}n_{+1}^{(2)}=\frac{k^{2}}{\omega^{2}}\phi_1^{(2)}+\frac{2ik(v_gk-\omega)}{\omega^{3}}\frac{\partial \phi_1^{(1)}}{\partial\xi},
\label{1:eq27}\\
&&\hspace*{-1.3cm}u_{+1}^{(2)}=\frac{k}{\omega}\phi_1^{(2)}+\frac{i(v_gk-\omega)}{\omega^{2}}\frac{\partial \phi_1^{(1)}}{\partial\xi},
\label{1:eq28}\\
&&\hspace*{-1.3cm}n_{-1}^{(2)}=-\frac{s_1k^{2}}{\omega^{2}}\phi_1^{(2)}-\frac{2ik s_1(v_gk-\omega)}{\omega^{3}}\frac{\partial \phi_1^{(1)}}{\partial\xi},
\label{1:eq29}\\
&&\hspace*{-1.3cm}u_{-1}^{(2)}=-\frac{k s_1}{\omega}\phi_1^{(2)}-\frac{i s_1(v_gk-\omega)}{\omega^{2}}\frac{\partial \phi_1^{(1)}}{\partial\xi},
\label{1:eq30}\
\end{eqnarray}
with compatibility condition
\begin{eqnarray}
&&\hspace*{-1.3cm}v_g=\frac{\partial \omega}{\partial k}=\frac{\omega(1+s_1 s_2-\omega^{2})}{k(1+s_1 s_2)}.
\label{1:eq31}\
\end{eqnarray}
The coefficients of $\epsilon$ for $m=2$ with $l=2$
provide the second-order harmonic amplitudes which are found to be
proportional to $|\phi_1^{(1)}|^{2}$
\begin{eqnarray}
&&\hspace*{-1.3cm}n_{+2}^{(2)}=M_4|\phi_1^{(1)}|^2,
\label{1eq:32}\\
&&\hspace*{-1.3cm}u_{+2}^{(2)}=M_5 |\phi_1^{(1)}|^2,
\label{1eq:33}\\
&&\hspace*{-1.3cm}n_{-2}^{(2)}=M_6|\phi_1^{(1)}|^2,
\label{1eq:34}\\
&&\hspace*{-1.3cm}u_{-2}^{(2)}=M_7 |\phi_1^{(1)}|^2,
\label{1eq:35}\\
&&\hspace*{-1.3cm}\phi_{2}^{(2)}=M_8 |\phi_1^{(1)}|^2,
\label{1eq:36}\
\end{eqnarray}
where
\begin{eqnarray}
&&\hspace*{-1.3cm}M_4=\frac{3 k^{4}+2M_8\omega^{2}k^{2}}{2\omega^{4}},
\nonumber\\
&&\hspace*{-1.3cm}M_5=\frac{k^{3}+2M_8k\omega^{2}}{2\omega^{3}},
\nonumber\\
&&\hspace*{-1.3cm}M_6=\frac{3 s_1^{2}k^{4}-2M_8s_1k^{2}\omega^{2}}{2\omega^{4}},
\nonumber\\
&&\hspace*{-1.3cm}M_7=\frac{s_1^{2}k^{3}-2ks_1M_8\omega^{2}}{2\omega^{3}},
\nonumber\\
&&\hspace*{-1.3cm}M_8=\frac{2M_2\omega^{4}+3s_2s_1^{2}k^{4}-3k^{4}}{2\omega^{2}(k^{2}+s_1s_2k^{2}-M_1\omega^{2}-4k^{2}\omega^{2})}.
\nonumber\
\end{eqnarray}
Now, we consider the expression for ($m=3$ with $l=0$) and ($m=2$ with $l=0$)
which leads the zeroth harmonic modes.
\begin{figure}[t!]
\centering
\includegraphics[width=80mm]{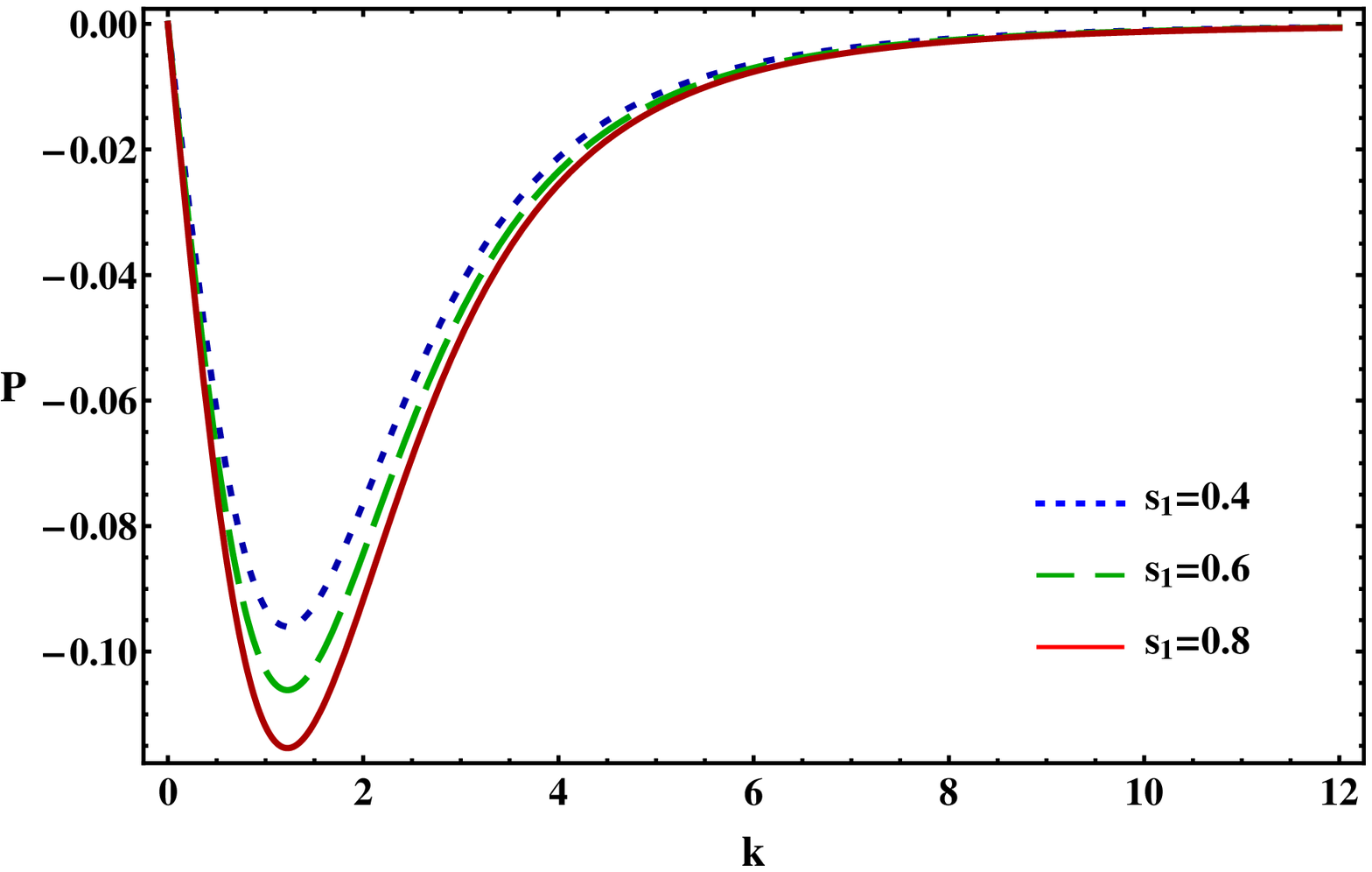}
\caption{Plot of $P$ vs $k$ for  various values of $s_1$ when $s_2=2.0$ and $\kappa=1.7$.}
\label{1Fig:F1}
\vspace{0.8cm}
\includegraphics[width=80mm]{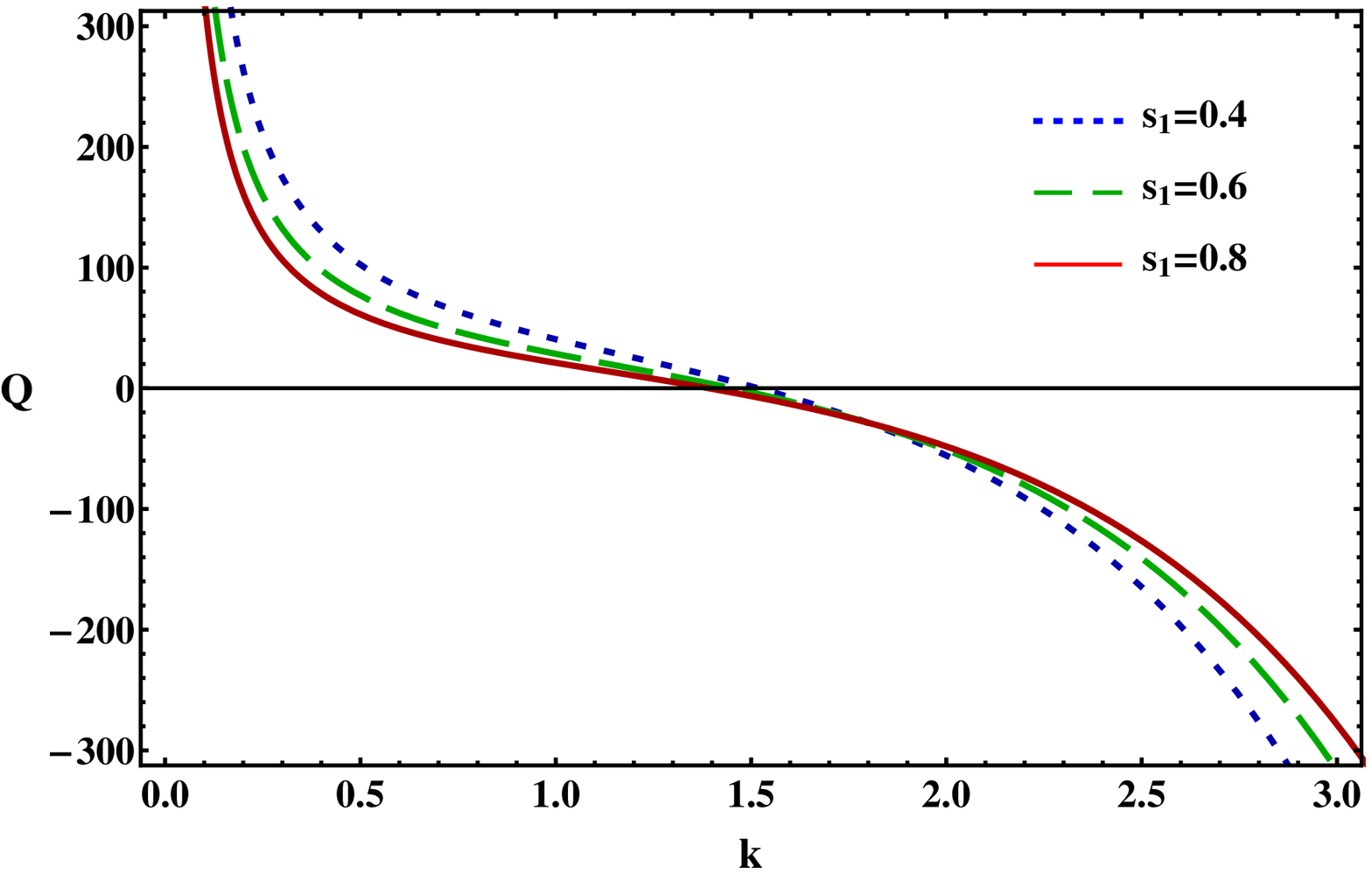}
\caption{Plot of $Q$ vs $k$ for various values of $s_1$ when $s_2=2.0$ and $\kappa=1.7$.}
 \label{1Fig:F2}
\end{figure}
Thus, we obtain
\begin{eqnarray}
&&\hspace*{-1.3cm}n_{+0}^{(2)}=M_{9}|\phi_1^{(1)}|^2,
\label{1:eq37}\\
&&\hspace*{-1.3cm}u_{+0}^{(2)}=M_{10}|\phi_1^{(1)}|^2,
\label{1:eq38}\\
&&\hspace*{-1.3cm}n_{-0}^{(2)}=M_{11}|\phi_1^{(1)}|^2,
\label{1:eq39}\\
&&\hspace*{-1.3cm}u_{-0}^{(2)}=M_{12}|\phi_1^{(1)}|^2,
\label{1:eq40}\\
&&\hspace*{-1.3cm}\phi_0^{(2)}=M_{13}|\phi_1^{(1)}|^2,
\label{1:eq41}\
\end{eqnarray}
where
\begin{eqnarray}
&&\hspace*{-1.3cm}M_9=\frac{2v_g k^{3}+\omega k^{2}+M_{13}v_g\omega^{3}}{v_g^{2}\omega^{3}},
\nonumber\\
&&\hspace*{-1.3cm}M_{10}=\frac{k^{2}+M_{13}\omega^{2}}{v_g\omega^{2}},
\nonumber\\
&&\hspace*{-1.3cm}M_{11}=\frac{2v_gs_1^{2}k^{3}+\omega k^{2}s_1^{2}-s_1 M_{13}\omega^{3}}{v_g^{2}\omega^{3}},
\nonumber\\
&&\hspace*{-1.3cm}M_{12}=\frac{k^{2}s_1^{2}-s_1 M_{13}\omega^{2}}{v_g\omega^{2}},
\nonumber\\
&&\hspace*{-1.3cm}M_{13}=\frac{2M_2v_g^{2}\omega^{3}-2v_g k^{3}-\omega k^{2}+2s_2 v_g s_1^{2}k^{3}+\omega s_2 k^{2}s_1^{2}}{\omega^{3}(1+s_1 s_2-M_1 v_g^{2})}.
\nonumber\
\end{eqnarray}
Finally, the third harmonic modes ($m=3$) and ($l=1$), with the help of \eqref{1:eq22}$-$\eqref{1:eq41},
give a set of equations which can be reduced to the
following NLSE:
\begin{eqnarray}
&&\hspace*{-1.3cm}i \frac{\partial\Phi}{\partial\tau} + P \frac{\partial^2\Phi}{\partial\xi^2}+ Q\Phi|\Phi|^2 =0,
\label{1:eq42}\
\end{eqnarray}
where $\Phi=\phi_1^{(1)}$ is used for simplicity. The dispersion coefficient $P$ is
\begin{eqnarray}
&&\hspace*{-1.3cm}P=\frac{3v_g(v_g k-\omega)}{2k\omega},
\nonumber\
\end{eqnarray}
and the  nonlinear coefficient $Q$ is
\begin{eqnarray}
&&\hspace*{-1.3cm}Q=\frac{2M_2\omega^3(M_8+M_{13})+3M_3\omega^3-R}{2k^{2}(1+s_1 s_2)},
\nonumber\
\end{eqnarray}
where
\begin{eqnarray}
&&\hspace*{-1.3cm}R=2k^{3}(M_5+M_{10})+2s_1 s_2k^{3}(M_7+M_{12})
\nonumber\\
&&\hspace*{+1.2cm}+\omega k^{2}(M_4+M_9)+s_1s_2\omega k^{2}(M_6+M_{11}).
\nonumber\
\end{eqnarray}
It may be noted here that both $P$ and $Q$ are function of various
plasma parameters such as $k$, $s_1$, $s_2$, and $\kappa$.
So, all the plasma parameters are used to maintain
the nonlinearity and the dispersion properties of the EDDPM.
\begin{figure}[t!]
\centering
\includegraphics[width=80mm]{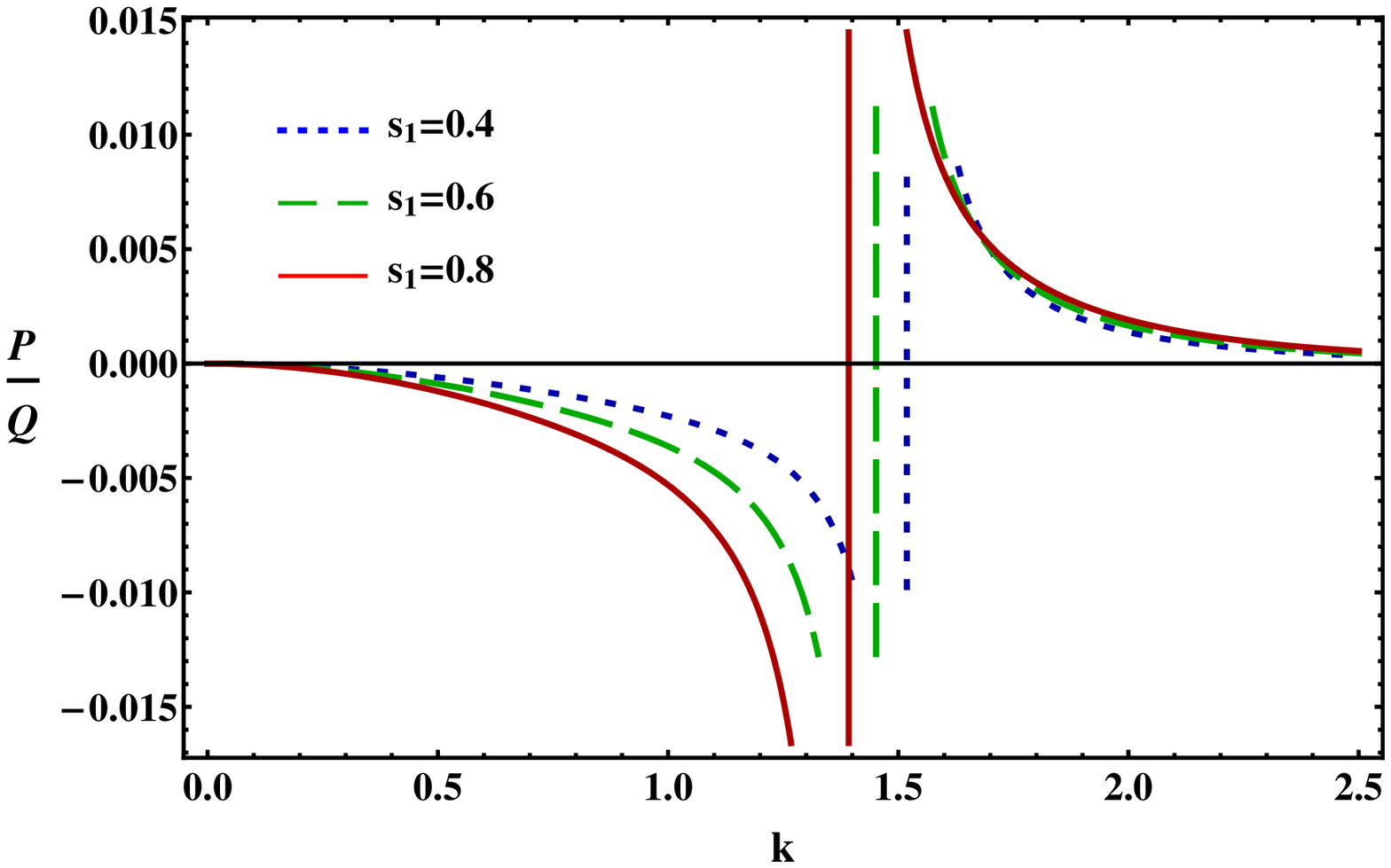}
\caption{Plot of $P/Q$ vs $k$ for various values of $s_1$ when $s_2=2.0$ and $\kappa=1.7$.}
\label{1Fig:F3}
\vspace{0.8cm}
\includegraphics[width=80mm]{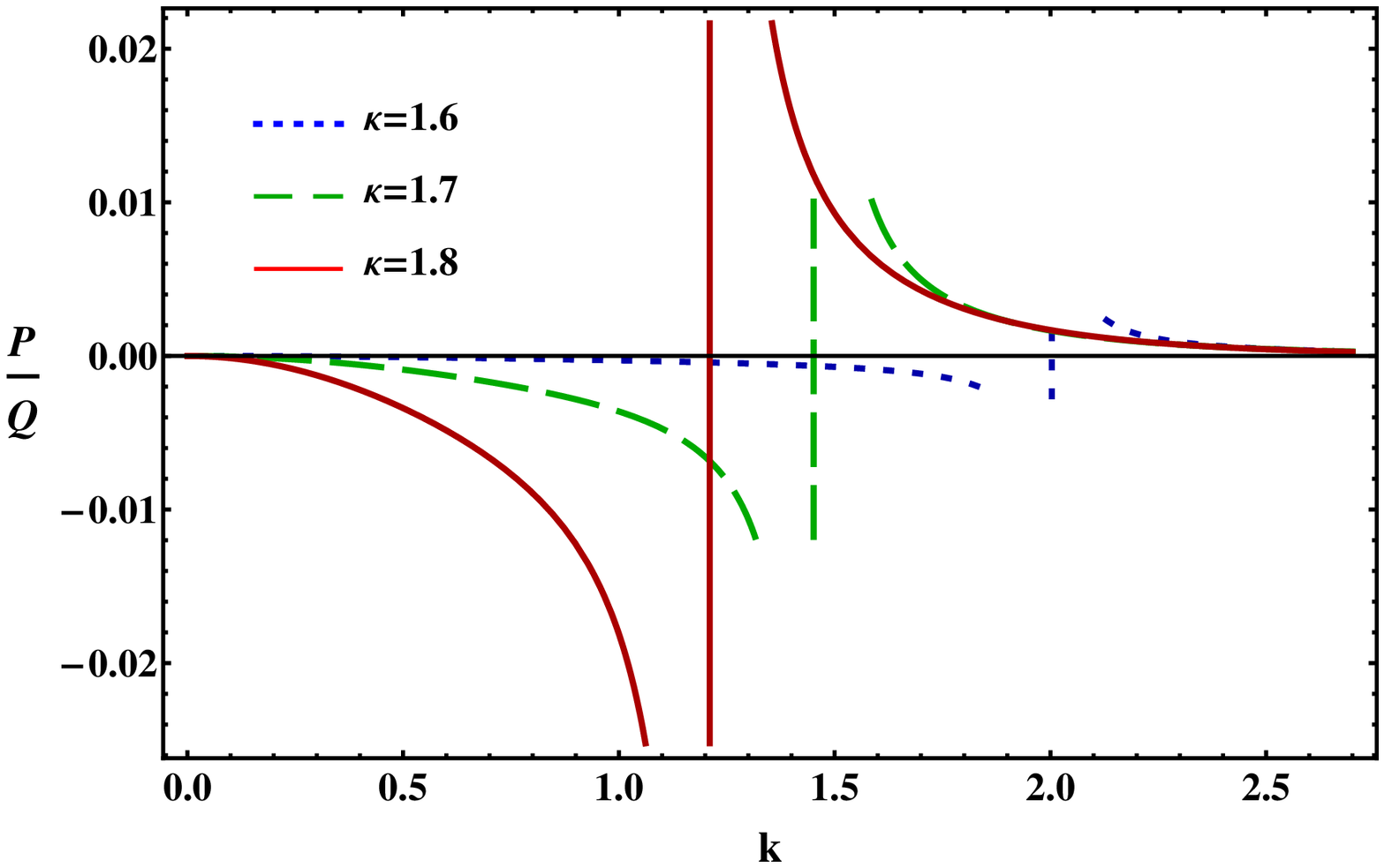}
\caption{Plot of $P/Q$ vs $k$ for various values of $\kappa$ when $s_1=0.6$ and $s_2=2.0$.}
 \label{1Fig:F4}
\end{figure}
\begin{figure}[t!]
\centering
\includegraphics[width=80mm]{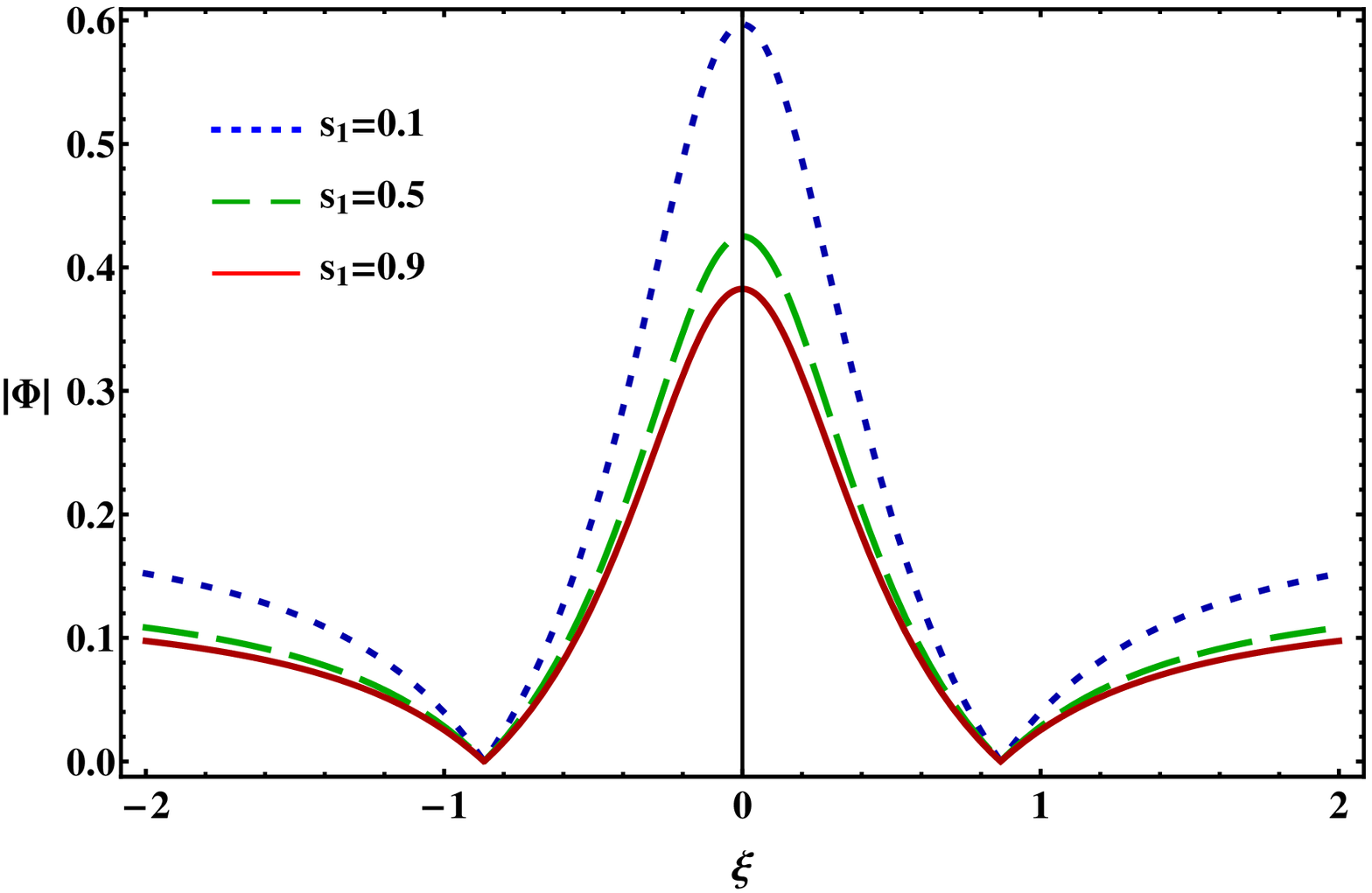}
\caption{Plot of $|\Phi|$ vs $\xi$ for various values of $s_1$ when  $k=1.6$, $\tau=0$, $s_2=2.0$, and $\kappa=1.7$.}
\label{1Fig:F5}
\vspace{0.8cm}
\includegraphics[width=80mm]{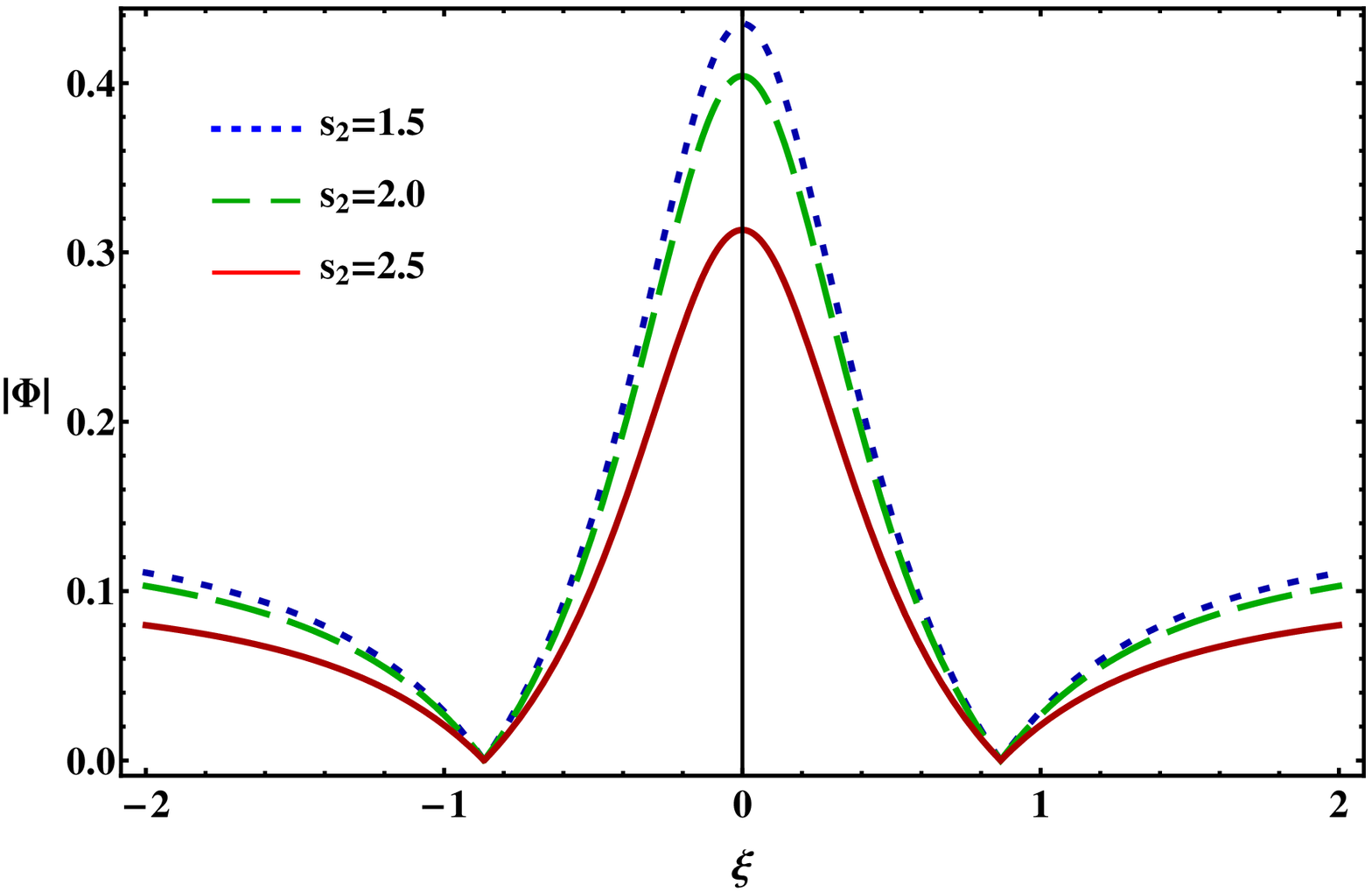}
\caption{Plot of $|\Phi|$ vs $\xi$ for various values of $s_2$ when $k=1.6$, $\tau=0$, $s_1=0.6$, and $\kappa=1.7$.}
 \label{1Fig:F6}
\end{figure}
\section{Modulational instability and Rogue waves}
\label{1:MI analysis and Rogue waves}
The stable and unstable parametric regimes of the DAWs are organized by the sign of the dispersion ($P$)
and nonlinear ($Q$) coefficients of the standard NLSE \eqref{1:eq42} \cite{Sultana2011,Ahmed2018,Gill2010,Saini2018,Kourakis2005}.
When $P$ and $Q$ are same sign (i.e., $P/Q>0$),
the evolution of the DAWs amplitude is modulationally unstable. On the other hand, when $P$ and $Q$ are
opposite sign (i.e., $P/Q<0$), the DAWs are modulationally stable in presence of the external perturbations.
The plot of $P/Q$ against $k$ yields stable and unstable domains for the DAWs.
The point, at which transition of $P/Q$ curve intersect with $k$-axis, is known as threshold
or critical wave number $k$ ($=k_c$).

The governing equation regarding the electron depleted DARWs in the modulationally
unstable parametric regime ($P/Q>0$) can be written as \cite{Akhmediev2009a,Ankiewiez2009b}
\begin{eqnarray}
&&\hspace*{-1.3cm}\Phi(\xi,\tau)= \sqrt{\frac{2P}{Q}} \left[\frac{4(1+4iP\tau)}{1+16P^2\tau^2+4\xi^2}-1 \right]\mbox{exp}(2iP\tau).
\label{1:eq43}
\end{eqnarray}
The plot of $P/Q$ vs $k$ for different plasma parameters can demonstrate the stable and unstable parametric regimes of DAWs.
In the unstable parametric regime DARWs are formed in an EDDPM due to the interaction of OPDGs with ions.
\section{Results and Discussion}
\label{1:Results and Discussion}
The stability conditions of the DAWs and the formation of the DARWs can be observed from Figs. \ref{1Fig:F1} to \ref{1Fig:F6}.
It is, however, clear from Fig. \ref{1Fig:F1} that (a) the $P$ is always negative for all positive values of $k$;
(b) the absolute value of the $P$ increases with increasing $s_1$, i.e., charge state of the negative dust
grain or mass of the positive dust grain when the charge state of the positive dust and the mass of the
negative dust grains remain constant. On the other hand, from Fig. \ref{1Fig:F2}, it can be manifested that (a) $Q$ is positive or
negative according to the values of $k$ and $s_1$, when other plasma parameters, namely, $s_2$ and $\kappa$ remain unchanged;
(b) $Q$ is positive (negative) for small (large) values of $k$.
This indicates that the instability criterion of the DAWs as well as generation of
the highly energetic DARWs in an EDDPM only to be determined by the sign of $Q$.

Figures \ref{1Fig:F3} and \ref{1Fig:F4} show two parametric regimes, one corresponding to the stable
(i.e., $P/Q<0$) DAWs and other corresponding to the unstable (i.e., $P/Q>0$ and indicating the
formation of the DARWs) DAWs in an EDDPM. These two parametric regimes, however, are separated by
a vertical line, and corresponding wave number is
known as critical wave number ($=k_c$) in ``$P/Q$  versus $k$" curve. The effects of positive and
negative dust masses and their charge states in recognizing the stable and unstable parametric regimes
associated with DAWs in an EDDPM can be observed from Fig. \ref{1Fig:F3}, and it is obvious from this
figure that (a) the $k_c$ decreases (increases) with an increase in the value of positive (negative) dust mass for
constant value of  negative and positive dust charge states; (b) on the contrary, the $k_c$
increases (decreases) with an increase in the value of $Z_+$ ($Z_-$) for constant value
of negative and positive dust masses (via $s_1$). So, the mass and charge state of the
positive and negative dust play an opposite role in recognizing the stability of the
DAWs in an EDDPM.

To examine the effects of the super-thermality of the positive ions to establish the
stable and unstable parametric regimes for DAWs in an EDDPM, we have depicted
Fig. \ref{1Fig:F4} and this figure indicates that (a) both stable (i.e., $P/Q<0$) and
unstable (i.e., $P/Q>0$ and indicating the formation of the DARWs) parametric regime
for DAWs can exist; (b) when $\kappa=1.6$, $1.7$, and $1.8$ then the
corresponding $k_c$ value is $k_c\equiv2.0$ (dotted blue curve), $k_c\equiv1.5$ (dashed green curve),
and $k_c\equiv1.2$ (solid red curve); (c) so, the $\kappa$ reduces the critical value hence the stable domain for the DAWs.

We have numerically analyzed Eq. \eqref{1:eq43} in Figs. \ref{1Fig:F5} and \ref{1Fig:F6} to understand how various plasma parameters
influence the nonlinearity as well as the formation of DARWs associated with unstable parametric regime of DAWs in an EDDPM.
The transformation of the amplitude of the carrier waves in a nonlinear dispersive medium is highly influenced by
the existence of OPDGs and their intrinsic properties (viz., charge and mass) as they
interfere with each other to organize nonlinear property, which describes the structure of the
DARWs associated with DAWs in the modulationally unstable parametric regime, of the EDDPM in presence
of the super-thermal ions can be seen from Figure \ref{1Fig:F5}, and it is clear from this figure that
the nonlinearity as well as the height and thickness of the DARWs in an EDDPM having super-thermal ions increases (decreases)
with increasing $Z_+$ ($Z_-$) for fixed value of $m_-$ and $m_+$ (via $s_1$).
The exact nature of the electrostatic DARWs according to the number density and charge
state of the OPDGs (via $s_2$) can be observed  from Fig. \ref{1Fig:F6},
and this figure exhibits that (a) the number density of negative (positive) dust in EDDPM  minimizes
(maximizes) the nonlinearity, i.e., the height as well as thickness of the DARWs in an EDDPM decreases
(increases) in space evolution for a constant value of time as well as negative and positive
dust charge states.
\section{Conclusion}
\label{1:Conclusion}
In this paper, we have emphasized not only the nonlinear and dispersive features of a
three component EDDPM but also the stability of the DAWs and construction of DARWs by
deriving a standard NLSE. The nonlinear and the dispersive coefficients of the standard
NLSE reflect the stable and unstable parametric regimes of the DAWs as well as the
mechanism to establish the gigantic DARWs associated DAWs in the unstable parametric
regime. The numerical analysis shows that the super-thermal ions have the capability to
control the MI of DAWs in an EDDPM, and also expresses that the MI conditions of the
DAWs in an EDDPM are also function of the intrinsic properties (viz.,
charge, mass, and number density) of the massive OPDGs as well as ions.
We can expect that the outcomes of our current work can be applicable in maximizing our
knowledge regarding the formation of the DARWs in EDDPM which are quite connected with
various space plasma, viz., the Earth polar mesosphere \cite{Hossen2016a}, interstellar space \cite{Hossen2016b},
cometary tails, Jupiter's magnetosphere, F-rings of Saturn \cite{Mayout2012}, and also laboratory plasma namely,
laser-matter plasma interaction \cite{Shahmansouri2013}.
\section*{Acknowledgment}
R. K. Shikha is thankful to the Bangladesh Ministry of Science and Technology
for awarding the National Science and Technology (NST) Fellowship.
A. Mannan gratefully acknowledges the financial support of the Alexander von
Humboldt-Stiftung (Bonn, Germany).


\begin{thebibliography}{99}
\bibitem{Shukla1992} P. K. Shukla and V. P. Silin, Phys. Scr. \textbf{45}, 508 (1992).

\bibitem{Hossen2016a} M. M. Hossen, M. S. Alam, S. Sultana, and A. A. Mamun, Eur. Phys. J. D \textbf{70}, 252 (2016).

\bibitem{Hossen2016b} M. M. Hossen, M. S. Alam, S. Sultana, and A. A. Mamun, Phys. Plasmas \textbf{23}, 023703 (2016).

\bibitem{Hossen2017} M. M. Hossen, L. Nahar, M. S. Alam, S. Sultana, and A. A. Mamun, High Energ. Dens. Phys. \textbf{24}, 9 (2017).

\bibitem{Shahmansouri2013} M. Shahmansouri and H. Alinejad, Phys. Plasmas \textbf{20}, 033704 (2013).

\bibitem{C1} M. H. Rahman, N. A. Chowdhury, A. Mannan, M. Rahman, and A. A. Mamun, Chinese J. Phys. \textbf{56}, 645 (2018).

\bibitem{C2} N. A. Chowdhury, A. Mannan, and A. A. Mamun, Phys. Plasmas  \textbf{24}, 113701 (2017).

\bibitem{C3} S. Jahan, N. A. Chowdhury, A. Mannan, and A. A. Mamun, Commun. Theor. Phys. \textbf{71}, 327 (2019).

\bibitem{Rao1990} N. N. Rao, P. K. Shukla, and M. Y. Yu, Planet. Space Sci. \textbf{38}, 543 (1990).

\bibitem{Barkan1995} A. Barkan, R. L. Merlino, and N.\'{D}. Angelo, Phys. Plasmas \textbf{2}, 3563 (1995).

\bibitem{Melandso1996} F. Melandso, Phys. Plasmas \textbf{3}, 3890 (1996).

\bibitem{Shukla1991} P. K. Shukla, M. Yu, and Y. R. Bharuthram, J. Geophys. Res. \textbf{96}, 21343 (1991).

\bibitem{Ferdousi2015} M. Ferdousi, M. R. Miah, S. Sultana, and A. A. Mamun, Astrophys. Space Sci. \textbf{43}, 360 (2015).

\bibitem{Mamun1996} A. A. Mamun, R. A. Cairns, and P. K. Shukla, Phys. Plasmas \textbf{3}, 702 (1996).

\bibitem{Dialynas2009} K. Dialynas, S. M. Krimigis, D. G. Mitchemm, D. C. Hamilton, N. Krupp, and P. C. Brandt, J. Geophys. Res. \textbf{114}, A01212 (2009).

\bibitem{Mayout2012} S. Mayout and M. Tribeche, J. Plasma Phys. \textbf{78}, 657 (2012).

\bibitem{Sahu2012} B. Sahu and M. Tribeche, Astrophys. Space Sci. \textbf{341}, 573 (2012).

\bibitem{Ferdousi2017} M. Ferdousi, S. Sultana, M. M. Hossen, M. R. Miah, and A. A. Mamun, Eur. Phys. J. D. \textbf{71}, 102 (2017).

\bibitem{Vasyliunas1968} V. M. Vasyliunas, J. Geophys. Res. \textbf{73}, 2839 (1968).

\bibitem{C4} M. H. Rahman, A. Mannan, N. A. Chowdhury, and A. A. Mamun, Phys. Plasmas \textbf{25}, 102118 (2018).

\bibitem{C5} N. A. Chowdhury, A. Mannan, M. M. Hasan, and A. A. Mamun, Plasma Phys. Rep. \textbf{45}, 459 (2019).

\bibitem{Shahmansouri2012} M. Shahmansouri and H. Alinejad, Phys. Plasmas \textbf{19}, 123701 (2012).

\bibitem{Kourakis2011} I. Kourakis and S. Sultana,  AIP Conf. Proc. \textbf{1397}, 86 (2011).

\bibitem{Uddin2015} M. J. Uddin, M. S. Alam, and A. A. Mamun, Phys. Plasmas \textbf{22}, 062111 (2015).

\bibitem{C6} N. A. Chowdhury, A. Mannan, M. R. Hossen, and A. A. Mamun, Contrib. Plasma Phys. \textbf{58}, 870 (2018).

\bibitem{Sultana2011} S. Sultana and I. Kourakis, Plasma Phys. Control. Fusion \textbf{53}, 045003 (2011).

\bibitem{Ahmed2018} N. Ahmed, A. Mannan, N. A. Chowdhury, and A. A. Mamun, Chaos \textbf{28}, 123107 (2018).

\bibitem{Gill2010} T. S. Gill, A. S. Bains, and C. Bedi, Phys. Plasmas \textbf{17}, 013701 (2010).

\bibitem{C7} N. A. Chowdhury, A. Mannan, M. M. Hasan, and A. A. Mamun, Chaos \textbf{27}, 093105 (2017).

\bibitem{Saini2018} N. S. Saini and I. Kourakis, Phys. Plasmas \textbf{15}, 123701 (2018).

\bibitem{C8} N. A. Chowdhury, M. M. Hasan, A. Mannan, and A. A. Mamun, Vacuum \textbf{147}, 31 (2018).

\bibitem{Kourakis2005} I. Kourakis and P. K. Shukla, J. Plasma Phys. \textbf{71}, 185 (2005).

\bibitem{Akhmediev2009a} N. Akhmediev, A. Ankiewicz, and J. M. Soto-Crespo, Phys. Rev. E \textbf{80}, 026601 (2009).

\bibitem{Ankiewiez2009b} A. Ankiewicz, N. Devine, and N. Akhmediev, Phys. Lett. A \textbf{373}, 3997 (2009).

\end{thebibliography}
\end{document}